\documentclass[aps,prl,preprint,showpacs,floatfix,preprintnumbers,amsmath,amssymb,superscriptaddress]{revtex4-1}
\usepackage{graphics}
\usepackage{graphicx}
\usepackage{epsfig}
\usepackage{amsmath}
\usepackage{bm}
\begin{document}
\title{Resonance in Magnetostatically Coupled Transverse Domain Walls}
\author{A. T. Galkiewicz}
\affiliation{School of Physics and Astronomy, University of Minnesota, 116 Church St.
SE, Minneapolis, MN 55455}
\author{L. O'Brien}
\affiliation{Department of Chemical Engineering and Materials Science, University of
Minnesota, 421 Washington Ave. SE, Minneapolis, MN 55455} \affiliation{Thin Film
Magnetism Group, Cavendish Laboratory, University of Cambridge, JJ Thomson Ave,
Cambridge CB3 OHE, United Kingdom}
\author{P. S. Keatley}
\affiliation{School of Physics, University of Exeter, Stocker Road, Exeter EX4 4QL,
United Kingdom}
\author{R. P. Cowburn}
\affiliation{Thin Film Magnetism Group, Cavendish Laboratory, University of Cambridge,
JJ Thomson Ave, Cambridge CB3 OHE, United Kingdom}
\author{P. A. Crowell}
\email{crowell@physics.umn.edu} \affiliation{School of Physics and Astronomy,
University of Minnesota, 116 Church St. SE, Minneapolis, MN 55455}
\begin{abstract}
We have observed the eigenmodes of coupled transverse domain walls in a pair of
ferromagnetic nanowires.  Although the pair is coupled magnetostatically, its spectrum
is determined by a combination of pinning by edge roughness and dipolar coupling of the
two walls. Because the corresponding energy scales are comparable, the coupling can be
observed only at the smallest wire separations.  A model of the coupled wall dynamics
reproduces the experiment quantitatively, allowing for comparisons with the estimated
pinning and domain wall coupling energies.  The results have significant implications
for the dynamics of devices based on coupled domain walls.
\end{abstract}
\pacs{75.60.Ch, 75.75-c, 75.78-n, 85.70.Kh}
\maketitle
\begin{figure}
\centerline{\epsfbox{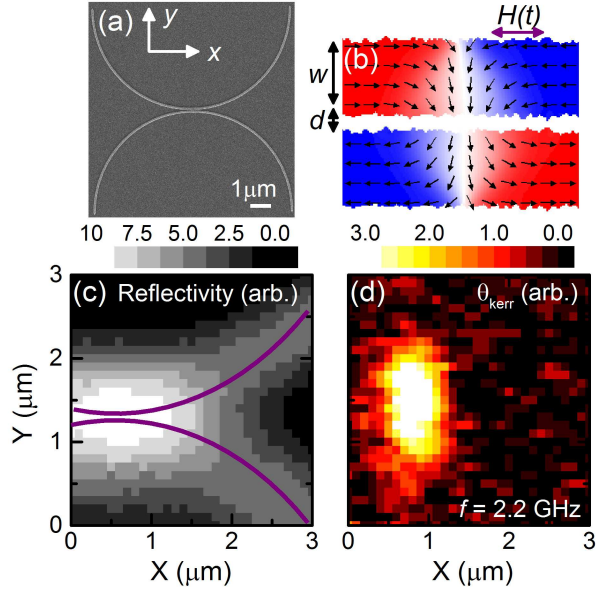}}
\caption{(color online) (a) Scanning electron micrograph of a $w=85$ nm, $d=40$ nm
device. (b) Micromagnetic simulation of two coupled DWs in a $w=100$ nm, $d=20$ nm pair
of nanowires. Arrows indicate the in-plane direction of $\mathbf{M}$, and colors the
magnitude of $M_x$ (+$M_s$ red, -$M_s$ blue). (c) 2D optical reflectivity map and (d)
Kerr rotation map of a $w=85$ nm, $d=130$ nm device. The solid lines in (c) are the
nanowire positions. The Kerr rotation map is shown for the maximum response on
resonance.}
\label{fig:fig1}
\end{figure}

Interactions between nano-scale ferromagnetic domains or domain walls (DWs) can become
significant as separations approach the dimensions of the individual structures. These
interactions can modify existing resonant behavior
\cite{Kuhlmann_Magnetization,Swoboda_Dynamic}, lead to the appearance of new coupled
modes \cite{Stamps_Domain,OBrien_Dynamic,Purnama_Current}, or result in a combination
of these \cite{Shibata_Dynamics,Jain_Coupled,Vogel_Coupled}. The coupling of DWs has
also been proposed as a scheme towards increasing the output power of spin-torque
oscillators \cite{Ruotolo_Phase}, and will have an impact on the design of
DW-logic-based devices as well. Understanding resonant behavior in these and similar
systems is thus of great interest, as resonances provide a direct probe of the
underlying coupling. The system of two interacting transverse domain walls (TDWs) in
parallel nanowires (NWs) [Fig. \ref{fig:fig1}(b)] is one of the simplest manifestations
of coupled DWs. Despite great interest in this particular coupling both experimentally
and numerically \cite{Purnama_Current,OBrien_Dynamic,OBrien_Near}, no experimental
observation of the coupled mode has yet been observed.

In this paper, we report on the observation of the resonant excitation of coupled TDWs
in adjacent NWs. By comparing such coupled excitations with those from single TDWs and
micromagnetic simulations, we show the experimental results may only be understood with
the inclusion of intrinsic pinning due to NW roughness: an effect predominantly
neglected when considering dynamic DW excitation. Extending simple one-dimensional (1D)
analytical models to describe phenomenologically both inter-DW coupling and intrinsic
DW pinning reproduces the observed spectra and provides insight into the interaction.
The frequency of one of the two modes of the coupled wall system is determined almost
entirely by the pinning energies of the individual walls, while only the second mode
depends significantly on the magnetotstatic coupling. We demonstrate that intrinsic
pinning and DW coupling remain comparable even at small inter-wire separations;
therefore pinning through roughness must always be considered when investigating such
dynamical DW experiments.

Figure \ref{fig:fig1}(a) shows a scanning electron microscope (SEM) image of the sample
geometry. Pairs of Permalloy semicircular NWs with radii of 5 $\mu$m and thickness of
10 nm were studied for NW widths $w$ of 70, 85, 140, and 190 nm, and closest
separations $d$ in the range of 40--140 nm. This design allows for repeatable DW
nucleation in the region of closest NW separation (here onwards referred to as the
interaction region) by the application of a saturation field along the $y$-direction
\cite{Saitoh_Current}. Based on the thickness and widths of the NWs, the stable
configuration is a TDW \cite{Laufenberg_Observation,Nakatani_Head}. This is also seen
in Fig. \ref{fig:fig1}(b) using micromagnetic simulations \cite{OOMMF} with a cell size
of $2\times 2\times 10$ $\text{nm}^3$, saturation magnetization $M_s=800$ kA/m,
exchange constant $A=13$ pJ/m, and Gilbert damping $\alpha=0.01$.

The basis of our measurement is time-resolved Kerr microscopy, which utilizes the
phase-locking of a microwave synthesizer (Agilent N5183A) to the 76 MHz repetition rate
of a pulsed Ti:Sapphire laser. The in-plane microwave field generated by the
synthesizer drives the motion of the DWs, which are probed stroboscopically by focusing
the pulse train (spot size of $\sim$400 nm) onto the interaction region using an oil
immersion objective. Through the use of the polar Kerr effect, we are sensitive to the
net out-of-pane component of the dynamic magnetization \cite{Chen_Nonlinear}. Figure
\ref{fig:fig1}(c) shows the spatial map of the sample reflectivity localized to the
interaction region of a $w=85$ nm, $d=130$ nm pair of NWs. The solid lines show the NW
positions. Figure \ref{fig:fig1}(d) --- acquired simultaneously with Fig.
\ref{fig:fig1}(c) --- shows the corresponding map of the $\theta_{Kerr}$ response for
an on-resonance excitation. Clearly the excited response is localized in the
interaction region, indicating that the observed spectra are due solely to the dynamic
response of the DWs.

We first investigate the response of a DW in a single NW. We expect, based on DW
propagation experiments, that pinning from intrinsic defects such as edge or surface
roughness will play a role in the dynamics \cite{Nakatani_Faster,Martinez_Thermal},
although this has not been thoroughly explored for resonant dynamics. For vortex domain
walls (VDWs) in NWs, studies have reported only on the characterization of a single
intrinsic \cite{Bedau_Quantitative} or patterned \cite{Moriya_Probe} pinning site.
Thus, it cannot be determined what influence intrinsic pinning (common to both cases)
has on the resonant frequency. A similar situation exists for TDWs in NWs
\cite{Bedau_Detection,Chang_Current}, however in this case an intrinsically pinned TDW
has not been investigated until now.

\begin{figure}
\centerline{\epsfbox{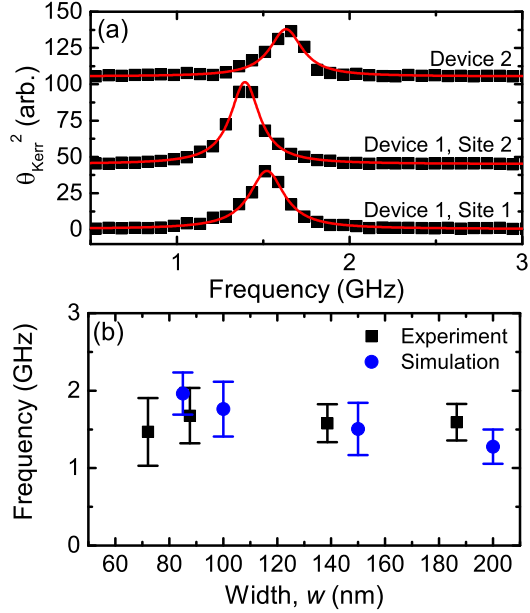}}
\caption{(color online) (a) DW spectra for $w=85$ nm nanowires taken at two DW
nucleation sites and in a separate device. Spectra are offset vertically for clarity.
The solid lines are Lorentzian fits from which the central frequency is extracted. (b)
Averaged pinning site frequency as a function of nanowire width for experiment
(squares) and micromagnetic simulation (circles). Error bars indicate dispersion of
pinned mode frequencies over the range of tested devices in experiment and simulation.}
\label{fig:fig2}
\end{figure}

Figure \ref{fig:fig2}(a) shows recorded spectra for isolated semicircular NWs that
reveal DW resonances in the range of 1--2 GHz. The spectra are obtained by plotting the
mean-squared amplitude of the time-resolved $M_z$ signals acquired for different
frequencies of the driving Oersted field. As shown, different resonant frequencies are
found when testing different NWs and when testing various DW nucleation sites in the
same NW. These observations are suggestive of a pinned mode (PM), in which the DW
oscillates in a local energy minimum created by the intrinsic roughness of the NW. The
variations in frequency in such a case are due to the random distribution of pinning
site depths and shapes. Figure \ref{fig:fig2}(b) shows the average PM frequency of four
devices at each width $w$. The error bars indicate the dispersion of observed
frequencies. In the tested range of $w$, we see no dependence of the PM frequency on
the width.

To confirm the origin of the PM, we perform dynamical micromagnetic simulations. A DW
is prepared in a simulated NW, and a 6 Oe, 100 ps wide Gaussian field pulse is applied
in the $x$-direction. The decay of the response is then recorded, and resonances are
found by taking the Fourier transform of the volume-averaged $M_z(t)$. Edge effects
from the simulation boundaries were determined to be negligible. We begin by simulating
an ideal NW having zero roughness. As expected, the DW moves freely along the NW and no
PM is observed. To include pinning, edge roughness, characterized by a root-mean-square
roughness $\sigma_{rms}$ and correlation length $\lambda$, is added to the NW. In these
simulations, a PM is observed that is a collective oscillation of the DW spins, and
results in the oscillating motion of the DW within the pinning site. From SEM images,
we estimate $\sigma_{rms}$ and $\lambda$ to be in the range of 1--4 nm and 2--6 nm
respectively. We find that the PM frequency is only weakly dependent on these two
parameters, due to an effective convolution of the edge roughness with the larger DW
profile ($\sim$100 nm). Figure \ref{fig:fig2}(b) shows the average PM frequency of six
simulated NWs versus width for $\sigma_{rms}=2$ nm and $\lambda=4$ nm, which match the
data well. A slight decrease in the PM frequency with width is observed in simulation,
which is the expected trend based on increases in the DW width and out-of-plane
anisotropy. We have also observed modes internal to the DW in simulation that are
characterized by a spatially non-uniform amplitude and phase \cite{Wang_Spin}, but due
to their nonuniformity we expect a weak coupling to the optical probe.

\begin{figure}
\centerline{\epsfbox{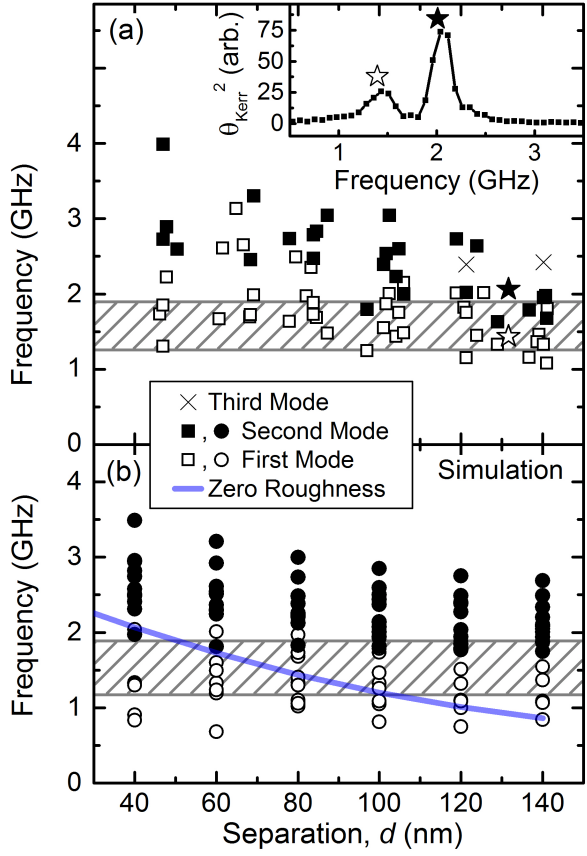}}
\caption{(color online) Separation dependence of (a) experimental and (b) simulated
coupled DW resonances. Open symbols correspond to the lowest frequency resonance
observed in each spectrum, closed symbols correspond to the next lowest frequency, and
a cross designates the two third modes seen in experiment. Data is included for all
nanowire widths. Hatched regions in (a) and (b) are the average plus or minus one
standard deviation of the single nanowire resonances from Fig. \ref{fig:fig2}. Inset in
(a) is the spectrum for a $w=85$ nm, $d=130$ nm nanowire pair with two resonances at
1.5 and 2 GHz. In (b) the solid line is the simulated DW-DW resonance for the case of a
$w=100$ nm pair of nanowires with zero edge roughness.}
\label{fig:fig3}
\end{figure}

We now turn our investigation to the case of coupled DWs. Resonant frequencies obtained
from the measured spectra are plotted in Fig. \ref{fig:fig3}(a) for different
separations $d$. As only slight ($<1$ GHz) width dependence is expected
\cite{OBrien_Dynamic} and no trend with respect to NW width is observed, we make no
distinction in $w$ in this plot. Of the 43 devices tested, 20 show two peaks in their
spectra and two devices show three peaks, the highest frequency of which are most
likely internal modes. The inset of Fig. \ref{fig:fig3}(a) is a sample spectrum that
shows two modes corresponding to the two starred points. Also included in this plot is
a hatched region that indicates the dispersion of PM frequencies from the single NW
case. While we may statistically correlate the first mode with a PM, the frequency of
the second mode is well above this region. In addition, the frequency of the second
mode decreases with increasing separation, as would be expected for a mode dependent on
the inter-DW coupling.

To further explore these modes, we utilize micromagnetics. We start by testing the case
of zero edge roughness. In contrast to the zero-roughness single NW simulations, a
translational mode is observed in the double NW system due to the inter-DW coupling.
Figure \ref{fig:fig3}(b) shows the results of these simulations for a $w=100$ nm pair,
plotted as a solid line. Comparing to Fig. \ref{fig:fig3}(a), the zero-roughness case
cannot explain the higher frequencies and the observation of multiple modes. We thus
proceed to test the system with the inclusion of edge roughness, using $\sigma_{rms}=2$
nm and $\lambda=4$ nm as previously characterized from the single NW data. In these
simulations the lowest two modes are translational, and plotting them gives the open
and closed circles in Fig. \ref{fig:fig3}(b), which shows good agreement with trends
observed in experiment. Simulations also show a qualitative similarity in the phase
response of the higher frequency mode compared to that of the zero-roughness DW-DW
mode.

\begin{figure}
\centerline{\epsfbox{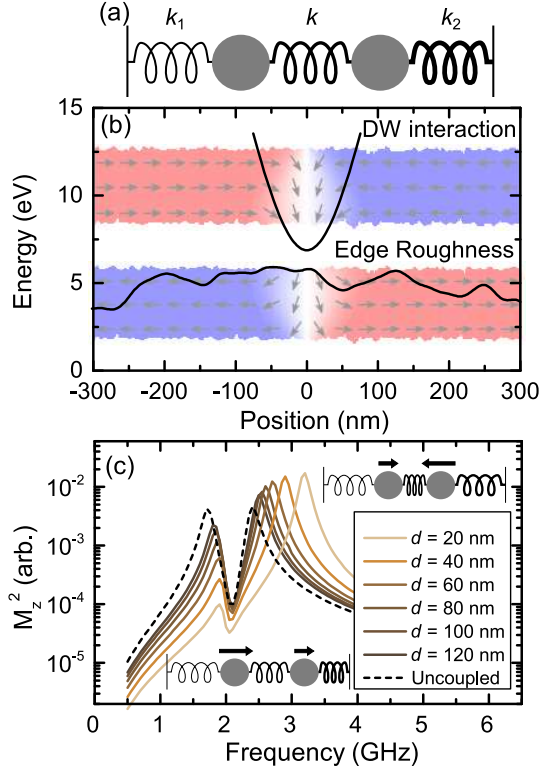}}
\caption{(color online) (a) Simplified model of the coupled DWs: two oscillating masses
coupled together with spring constant $k$. (b) Energy profiles for the edge roughness
of a $w=100$ nm nanowire (lower) and the DW interaction (upper) for a $w=100$ nm,
$d=20$ nm pair, as a function of the position of the DWs in the nanowires, offset for
clarity. Nanowires shown in the background share the x-axis scale. (c) Numerically
solved $M_z^2=(\sin\phi_1+\sin\phi_2)^2$ of the 1-D coupled DW equations of motion [Eq.
(\ref{eq:1Deom})] as a function of drive frequency and plotted for multiple nanowire
separations, shown for the case of two $w=100$ nm nanowires. Values used were
$k_1=0.21$ erg/$\text{cm}^2$, $k_2=0.40$ erg/$\text{cm}^2$, and $k=0.03-0.22$
erg/$\text{cm}^2$. Insets in bottom left (upper right) are representations of the lower
(higher) frequency mode.}
\label{fig:fig4}
\end{figure}

Given this quantitative agreement, we now look for an analytical understanding of this
system using the 1D equations of motion \cite{Thiaville_Domain,OBrien_Dynamic}. In
these equations the DW is described by $x$ and $\phi$, the center coordinate and angle
with respect to the $xy$ plane respectively. Letting $i$ = 1 or 2 denote the top or
bottom DW, the equations of motion are
\begin{subequations}
\begin{align}
Q_i\frac{\dot{x}_i}{\Delta}-\alpha\dot{\phi}_i &=\frac{\gamma}{M_s}K_s \sin(2\phi_i),
\\ Q_i\dot{\phi}_i+\frac{\alpha}{\Delta}\dot{x}_i &=\frac{\gamma}{2M_s
S}\left[-\frac{\partial U_i}{\partial x_i}\right].
\end{align}\label{eq:1Deom}%
\end{subequations}
Here $Q_1=1$, $Q_2=-1$ are the effective magnetostatic charges of the two DWs, $\Delta$
is the DW width, $K_s$ is the out-of-plane anisotropy, $\gamma$ is the gyromagnetic
ratio, and $S$ is the cross-sectional area of the nanowire. $U_i$ is the total energy
of a DW and is given by $U_i=U_0-Q_i M_s S H x_i+ k(x_1-x_2)^2+k_i x_i^2$, where the
terms on the right hand side are the internal energy, Zeeman energy, interaction
energy, and pinning energy due to roughness respectively. We have assumed a harmonic
potential for the roughness. In the limit of small $\phi_i$, Eq. (\ref{eq:1Deom}) can
be diagonalized and the natural frequencies of the resulting two modes are given by
\begin{equation}
\omega_{0}^{2}=\frac{k_1+k_2+2k}{m_D} \pm \frac{\sqrt{(k_1-k_2)^2+(2k)^2}}{m_D},
\label{eq:coupled_osc}
\end{equation}
where $m_D=(1+\alpha^2)M_s^2 S/\gamma^2 \Delta K_s$ is the DW D\"{o}ring mass
\cite{Doring}. Figure \ref{fig:fig4}(a) shows the simplified description of this system
as a coupled oscillator problem in which each mass (DW) sits in its own potential
(pinning site). We note here that an in-phase response of the DW displacements $x_i$ is
accompanied by an out-of-phase response of  $\phi_i$ and vice versa, due to the
coupling of $x$ and $\phi$ in Eq. (\ref{eq:1Deom}). We also note that given two
dissimilar pinning site spring constants, both modes of Eq. (\ref{eq:coupled_osc}) will
couple to the driving magnetic field and the relative oscillation amplitudes of the two
DWs will be unequal. This enables both the in-phase and out-of-phase modes to be
excited and detected experimentally. From Eq. (\ref{eq:coupled_osc}), as $k$ goes to
zero ($d\rightarrow\infty$) the two modes approach the respective PM frequencies. As
$k$ becomes large, the lower mode goes to the root-mean-square of the two PM
frequencies, while the upper mode goes to the expected DW-DW resonance. Estimates of
the spring constants can be found from energy landscapes calculated using
micromagnetics which are illustrated in Fig. \ref{fig:fig4}(b). To find the energy
profile due to edge roughness, we translate a DW profile through the length of a NW and
compute the total energy of the system at each position. We find the energy of the DW
coupling by separating the two DWs laterally in zero-roughness NWs and computing the
sum of the magnetostatic and exchange energies for different values of these
separations. Note that the ranges of the DW interaction potential and a given pinning
site are both set by the domain wall width.  This is why the curvatures of the
interaction potential and any given minimum of the magnetostatic potential due to edge
roughness in Fig. 4(b) are comparable, in spite of their very different depths.

Using the estimates for the $k_{i}$ and $k$ obtained from micromagnetics, we
numerically solve Eq. (\ref{eq:1Deom}) and plot the power spectrum of the net $M_z$
component for multiple $d$ [Fig. \ref{fig:fig4}(c)]. At large separations, the two DWs
tend to their independent PM frequencies and phases. As $d$ decreases, the lower mode,
illustrated in the bottom left inset, is suppressed due to the unfavorable driving
force and the out-of-phase response of the $M_z$ components. These observations can
explain the detection of only a single mode in some of the spectra: at large
separations two accidentally degenerate PM frequencies will be unresolvable, while at
small separations the lower frequency mode may be below the detection threshold.
Looking at the upper mode (depicted in the top right inset) we see an increase in
amplitude and frequency with decreasing $d$. The frequency dependence produced by this
model agrees strongly with the trends observed in both micromagnetics and experiment.
While experimental limitations such as non-ideal waveguides prevent a quantitative
comparison of the relative mode amplitudes, in general a qualitative agreement is also
seen [e.g. inset of Fig. \ref{fig:fig3}(a)].  From these observations, it is clear that
the first two modes in Fig. \ref{fig:fig3}(a) can be attributed to the two eigenmodes
of Eq. (\ref{eq:coupled_osc}), given a random distribution of pinning sites in the NWs.
As the width of this distribution is set in part by the DW itself, pinning will always
be significant in the dynamics, and must be considered in future experiments.

\begin{acknowledgments}
This work was supported in part by the NSF MRSEC program under DMR-0804244 and the
NSF/NRI NEB program under ECCS-1124831, as well as the EU Marie Curie IOF project no.
299376 and the European Community Seventh Framework Programme Contract No. 247368:
3SPIN. Parts of this work were carried out in the Characterization Facility, University
of Minnesota, which receives partial support from NSF through the MRSEC program.
\end{acknowledgments}


\begin{thebibliography}{99}

\bibitem{Kuhlmann_Magnetization}
N. Kuhlmann, A. Vogel, and G. Meier, Phys. Rev. B \textbf{85}, 014410 (2012)

\bibitem{Swoboda_Dynamic}
C. Swoboda, N. Kuhlmann, M. Martens, A. Vogel, and G. Meier, J. Appl. Phys.
\textbf{114}, 043905 (2013)

\bibitem{Stamps_Domain}
R. L. Stamps, A. S. Carri\c{c}o, and P. E. Wigen, Phys. Rev. B \textbf{55}, 6473 (1997)

\bibitem{OBrien_Dynamic}
L. O'Brien, E. R. Lewis, A. Fern\'andez-Pacheco, D. Petit, R.P. Cowburn, J. Sampaio,
and D. E. Read, Phys. Rev. Lett. \textbf{108}, 187202 (2012).

\bibitem{Purnama_Current}
I. Purnama, M. Chandra Sekhar, S. Goolaup, and W. S. Lew, Appl. Phys. Lett.
\textbf{99}, 152501 (2011).

\bibitem{Shibata_Dynamics}
J. Shibata, K. Shigeto, and Y. Otani, Phys. Rev. B \textbf{67}, 224404 (2003)

\bibitem{Jain_Coupled}
S. Jain, H. Schultheiss, O. Heinonen, F. Y. Fradin, J. E. Pearson, S. D. Bader, and V.
Novosad, Phys. Rev. B. \textbf{86}, 214418 (2012)

\bibitem{Vogel_Coupled}
A. Vogel, T. Kamionka, M. Martens, A. Drews, K. W. Chou, T. Tyliszczak, H. Stoll, B.
Van Waeyenberge, and G. Meier, Phys. Rev. Lett. \textbf{106}, 137201 (2011)

\bibitem{Ruotolo_Phase}
A. Ruotolo, V. Cros, B. Georges, A. Dussaux, J. Grollier, C. Deranlot, R. Guillemet, K.
Bouzehouane, S. Fusil, and A. Fert, Nat. Nanotechnol. \textbf{4}, 528 (2009)

\bibitem{OBrien_Near}
L. O'Brien, D. Petit, H. T. Zeng, E. R. Lewis, J. Sampaio, A. V. Jausovec, D. E. Read,
and R. P. Cowburn, Phys. Rev. Lett. \textbf{103}, 077206 (2009)

\bibitem{Saitoh_Current}
E. Saitoh, H. Miyajima, T. Yamaoka, and G. Tatara, Nature \textbf{432}, 203 (2004)

\bibitem{Laufenberg_Observation}
M. Laufenberg, D. Bedau, H. Ehrke, M. Kl\"{a}ui, U. R\"{u}diger, D. Backes, L. J.
Heyderman, F. Nolting, C. A. F. Vaz, J. A. C. Bland, T. Kasama, R. E. Dunin-Borkowski,
S. Cherifi, A. Locatelli, and S. Heun, Appl. Phys. Lett. \textbf{88}, 212510 (2006).

\bibitem{Nakatani_Head}
Y. Nakatani, A. Thiaville, and J. Miltat, J. Magn. Magn. Mater. \textbf{290}, 750
(2005).

\bibitem{OOMMF}
M. J. Donahue and D. G. Porter, Interagency Report NISTIR 6376 (National Institute of
Standards and Technology, Gaithersburg, MD, 1999).

\bibitem{Chen_Nonlinear}
T. Y. Chen and P. A. Crowell, IEEE Trans. Magn. \textbf{46}, 1457 (2010).

\bibitem{Nakatani_Faster}
Y. Nakatani, A. Thiaville, and J. Miltat, Nat. Mater. \textbf{2}, 521 (2003)

\bibitem{Martinez_Thermal}
E. Martinez, L. Lopez-Diaz, L. Torres, C. Tristan, and O. Alejos, Phys. Rev. B
\textbf{75}, 174409 (2007)

\bibitem{Bedau_Quantitative}
D. Bedau, M. Kl\"{a}ui, M. T. Hua, S. Krzyk, U. R\"{u}diger, G. Faini, and L. Vila,
Phys. Rev. Lett. \textbf{101}, 256602 (2008)

\bibitem{Moriya_Probe}
R. Moriya, L. Thomas, M. Hayashi, Y. B. Bazaliy, C. Rettner, and S. S. P. Parkin, Nat.
Phys. \textbf{4}, 368 (2008)

\bibitem{Bedau_Detection}
D. Bedau, M. Kl\"{a}ui, S. Krzyk, U. R\"{u}diger, G. Faini, and L. Vila, Phys. Rev.
Lett. \textbf{99}, 146601 (2007)

\bibitem{Chang_Current}
L. J. Chang, P. Lin, and S. F. Lee, Appl. Phys. Lett. \textbf{101}, 242404 (2012)

\bibitem{Wang_Spin}
X. Wang, G. Guo, G. Zhang, Y. Nie, Q. Xia, and Z. Li, J. Magn. Magn. Mater.
\textbf{332}, 56 (2013)

\bibitem{Thiaville_Domain}
A. Thiaville and N. Nakatani, in \emph{Spin Dynamics in Confined Magnetic Structures
III} (Springer, New York, 2006).

\bibitem{Doring}
W. Doring, Z. Naturforsch. A \textbf{3}, 373 (1948).


\end{thebibliography}
\end{document}